\documentstyle[aps,preprint]{revtex}
\begin{document}

\title{Coherent source radius in $p\bar{p}$ collisions}

\author{ Q. H. Zhang} 

\address{Institut f\"ur Theoretische Physik, Universit\"at 
Regensburg, D-93040 Regensburg, Germany}

\author{ X. Q. Li} 

\address{China Center of Advanced Science and Technology (World Laboratory),
P.O. Box 8730, Beijing 100080, P.R. China\\
Physics Department, Nankai University, Tianjing 300071, P.R. China }

\date{\today}

\maketitle

\begin{abstract}

We use a recently derived result to extract from two-pion interferometry data 
from $p\bar{p}$ collisions the radius of the coherent component in the source. We 
find a coherent source radius of about $2 fm$.

\end{abstract}

PACS numbers: 25.75. -q, 25.70. Pq, 25.70 Gh. 

Two-particle Bose-Einstein interferometry (also known as Hanbury 
Brown-Twiss intensity interferometry) as a method for obtaining 
information on the space-time geometry and dynamics of high energy 
collisions has recently received intensive theoretical and 
experimental attention. Detailed investigations have shown that high-quality 
two-particle correlation data can reveal not only the geometric 
extension of the particle-emitting source but also its dynamical state 
at particle freeze-out\cite{BGJ,APW,Pratt95,He96,Zajc86,Lorstad89}. 

For a partially coherent source, it was suggested that two-pion interferometry 
can give the coherent degree information of the source\cite{FW78,GKW79,Weiner89}. 
In paper\cite{He97},
we have generalized the two-pion interferometry formula for a partially 
coherent source and found that two-pion interferometry was also 
senstive to the size of the coherent source component.  The physical reason is clear: 
The particle spectrum distribution is the Fourier transformation 
of the source distribution, the coherent source radius information must hide in 
the spectrum distribution and also must appear in the two-pion interferometry 
formula. We will shown below  that a general two-pion interferometry formula 
satisfy the needs to reveal both the degree of coherence  and space-time information 
of the coherent source which was neglected in previous 
publications\cite{Weiner89,Alex93,Mark,UA1,KL90}.

For a partially coherent source, one can decompose the pion current 
into a coherent and a chaotic term\cite{GKW79,He97,CGZ}:
\begin{eqnarray}
J(x)=J_{c}(x)+J_{I}(x),
\end{eqnarray}
with 
\begin{eqnarray}
J_{I}(x)=\int d^4x' d^4 p j(x',p)\gamma(x')exp(ip(x-x')).
\end{eqnarray}
Here $j(x',p)$ is the probability amplitude of finding a pion 
with momentum $p$, emitted by the emitter at $x'$. 
$\gamma(x')=exp(i\phi(x'))$ is a 
random phase factor which has been taken away from $j(x',p)$. All 
emitters are uncorrelated when assuming:
\begin{eqnarray}
<\gamma^*(x)\gamma(y)>=\delta^4(x-y).
\end{eqnarray} 
Here $<\cdot \cdot \cdot>$ means phase average. Then the probability amplitude 
in momentum space reads:
\begin{eqnarray}
\tilde{J}(p)&=&\int J(x) exp(-ip\cdot x)d^4 x
\nonumber\\
&=&\tilde{J}_{I}(p)+\tilde{J}_{c}(p) ,
\end{eqnarray}
with
\begin{eqnarray}
\tilde{J}_{I}(p)=\int J_{I}(x) exp(-ip\cdot x) d^4x,~~ 
\tilde{J}_{c}(p)=\int J_{c}(x) exp(-ip\cdot x) d^4x.
\end{eqnarray}
Using the relationship
\begin{eqnarray}
<J^*_{I}(x)J_{c}(y)>=0,
\end{eqnarray}
we have 
\begin{eqnarray}
<\tilde{J}^*_{I}(p)\tilde{J}_{c}(p)>=0.
\end{eqnarray}
Then the single pion spectrum can be expressed as:
\begin{eqnarray}
P_1(p)&=&<\tilde{J}^*(p)\tilde{J}(p)>
\nonumber\\
&=&<\tilde{J}^*_{I}(p)\tilde{J}_{I}(p)>+\tilde{J}^*_{c}(p)\tilde{J}_{c}(p).
\end{eqnarray}
Using relationship 
\begin{eqnarray}
<\gamma^*(x_1)\gamma^*(y_1)\gamma(x_2)\gamma(y_2)>&=&
<\gamma^*(x_1)\gamma(x_2)><\gamma^*(y_1)\gamma(y_2)>
\nonumber\\
&&+<\gamma^*(x_1)\gamma(y_2)><\gamma^*(y_1)\gamma(x_2)>,
\end{eqnarray}
we obtain
\begin{eqnarray}
<\tilde{J}^*_{I}(p_1)\tilde{J}_{I}^*(p_2)\tilde{J}_{I}(p_2)\tilde{J}_{I}(p_1)>
=<|\tilde{J}_{I}(p_1)|^2>\cdot <|\tilde{J}_{I}(p_2)|^2>
+|<\tilde{J}_{I}^*(p_1)\tilde{J}_{I}(p_2)>|^2.
\end{eqnarray}
Performing the phase average and using Eq.(4), Eq.(8) and Eq.(10), we can 
write the two-pion spectrum as
\begin{eqnarray}
P_2(p_1,p_2)&=&<\tilde{J}^*(p_1)\tilde{J}^*(p_2)\tilde{J}(p_2)\tilde{J}(p_1)>
\nonumber\\
&=&P_1(p_1)P_1(p_2)+ |<\tilde{J}^*_{I}(p_1)\tilde{J}_{I}(p_2)>|^2
\nonumber\\
&&+ 2Re(<\tilde{J}^*_{I}(p_1)\tilde{J}_{I}(p_2)>\tilde{J}^*_{c}(p_2)\tilde{J}_{c}(p_1)).
\end{eqnarray}
Similar to ref.\cite{He97}, we define the following functions:
\begin{eqnarray}
g_{I}(x,k)=\int d^4 y <J_I^*(x+y/2)J_I(x-y/2)> \exp(ik\cdot y), 
\end{eqnarray}
\begin{eqnarray}
g_{c}(x,k)=\int d^4 y J_c^*(x+y/2)J_c(x-y/2) \exp(ik\cdot y). 
\end{eqnarray}
Here $g_I(x,p)$ and $g_c(x,p)$ are Wigner functions which describes the source 
distribution of the chaotic source and the coherent source respectively \cite{He97,Pratt84,CGZ}. 
From the above definitions, we have:
\begin{eqnarray}
\rho_I(i,j)=\rho_I(p_i,p_j)=<\tilde{J}^*_I(p_i)\tilde{J}_I (p_j)>
=\int g_I(x,\frac{p_i+p_j}{2}) \exp(i(p_i-p_j)\cdot x) d^4x
\end{eqnarray}
and 
\begin{eqnarray}
\rho_c(i,j)=\rho_c(p_i,p_j)=\tilde{J}^*_c(p_i)\tilde{J}_c(p_j)
=\int g_c(x,\frac{p_i+p_j}{2}) \exp(i(p_i-p_j)\cdot x) d^4x.
\end{eqnarray}
Then the two-pion correlation function for a partially coherent source can  be 
expressed as:
\begin{eqnarray}
C_2(p_1,p_2)=\frac{P_2(p_1,p_2)}{P_1(p_1)\cdot P_1(p_2)},
\end{eqnarray}
with
\begin{eqnarray}
P_2(p_1,p_2)&=P_1(p_1)\cdot P_1(p_2)+|\rho_I(1,2)|^2+2 Re(\rho_I(1,2)\rho_c(2,1))
\end{eqnarray}
and
\begin{eqnarray}
P_1(p)=\rho_I(p,p)+\rho_c(p,p).
\end{eqnarray}
For simplicity\cite{GKW79}, we assume that $\rho_I(i,j)$ and $\tilde{j}_c(p)$ are real,
then Eq. (16) can be re-written as\cite{Weiner89,GKW79} 
\begin{equation}
C_2(p_1,p_2)=1+\epsilon(p_1)\epsilon(p_2)H_I^2(p_1,p_2)+2\sqrt{\epsilon(p_1)
\epsilon(p_2)(1-\epsilon(p_1))(1-\epsilon(p_2))}H_I(p_1,p_2)
\end{equation}
with
\begin{eqnarray}
H_I(p_1,p_2)=\frac{\rho_I(1,2)}{\sqrt{\rho_I(p_1,p_1)\rho_I(p_2,p_2)}},
\epsilon(p)=\frac{\rho_I(p,p)}{P_1(p)}.
\end{eqnarray}
Here $\epsilon(p)$ describes the chaotic degree of the source. For a totally 
chaotic source,  $\epsilon(p)=1$ while for a totally coherent source,  $\epsilon(p)=0$. 
Due to the limited statistics,  the dependence of $\epsilon(p)$ on 
momentum $p$ is often neglected,  then the above formula can be 
simplified as\cite{Weiner89,Alex93,Mark}
\begin{eqnarray}
C_2(p_1,p_2)=1+\epsilon^2 H_I^2(p_1,p_2)+2\epsilon(1-\epsilon) H_I(p_1,p_2) .
\end{eqnarray}
Eq. (21) is now widely used in the literature. 
In the following, we will show that Eq. (21) is too simple to approximate 
Eq.(19) and the information about the radius of the coherent source is lost in Eq.(21). 
We will derive a formula which is better than Eq.(21) to 
approximate Eq. (19).  Eq. (16) can be re-written as 
\begin{equation}
C_2(p_1,p_2)=1+  \frac{|\rho_I(1,2)|^2}{P_1^2(k)}\cdot
\frac{P_1^2(k)}{P_1(p_1)P_1(p_2)}+
2Re\left( \frac{\rho_c(1,2)\rho_I(2,1)}{P_1^2(k)} \right)
\frac{P_1^2(k)}{P_1(p_1)P_1(p_2)}  .
\end{equation}
Here $k=(p_1 + p_2)/2 $.  Because the single particle distribution is 
exponential form in $p\bar{p}$ and $ee^+$ collisions\cite{He96},  we have 
$P_1^2(k)/P(p_1)/P(p_2)\sim 1$.  Under the assumption that 
$\rho_I(i,j)$ and $\tilde{j}_c(p)$ are real, the above formula can then be 
re-expressed as
\begin{eqnarray}
C(p_1,p_2)=1+\epsilon^2(k)R_I^2(p_1,p_2)+2\epsilon(k)(1-\epsilon(k))
R_c(2,1)R_I(1,2)
\end{eqnarray}
with
\begin{eqnarray}
R_I(p_1,p_2)=\frac{\rho_I(1,2)}{\rho_I(k,k)} ,
R_c(p_1,p_2)=\frac{\rho_c(1,2)}{\rho_c(k,k)}  .
\end{eqnarray}
One clearly sees the difference between Eq.(23) and Eq.(21), 
a extra term $R_c(p_1,p_2)$ appears in Eq.(23).
Similar to ref.\cite{KL90},  we take  
both $R_I(p_1,p_2)$ and $R_c(p_1,p_2)$ 
 as Gaussian functions and neglect the dependence of $\epsilon$ on 
momentum, then Eq.(23) can be written down as 
\begin{eqnarray}
C_2(p_1,p_2)=1+\epsilon^2 exp(-r_I^2\cdot q^2)+2\epsilon(1-\epsilon) 
exp(-(r_I^2+r_c^2)\cdot q^2/2) .
\end{eqnarray}
Here $r_I$ and $ r_c$ is the radius of the chaotic source and the coherent 
source 
respectively.  Eq.(25) is a limiting form  of the general expressions derived 
in Ref\cite{He97} for the case of a 1-dimensional Gaussian parametrization. 
In the above derivation, we did not take into account the influence of resonance decay 
on the two-pion correlations which 
may distort the simple Gaussian function assumed above and may also result in 
an effective double Gaussian two-pion correlation function.
 In paper\cite{KL90}, Kulka et al. used the following formula 
\begin{equation}
C_2(p_1,p_2)=1+\lambda_1\cdot exp(-R_1^2\cdot q^2)+
\lambda_2 exp(-R_2^2\cdot q^2)
\end{equation}
to fit their data and found that 
$\lambda_1=0.30^{+0.08}_{-0.09}$, $\lambda_2=0.45 \pm 0.13$, $R_1=0.63 \pm 0.04 fm$ and 
$R_2=2.11\pm 0.63 fm$ with $\chi^2/NDF=1.09$. In the paper
\cite{KL90}, Kulka et al found that Eq.(26) was not consistent with Eq. (21). 
 But if we take those values into Eq.(25), we will 
find that $r_I=0.63 \pm 0.04 ~fm$ ,$r_c=2.01 \pm 0.67 ~fm$ and $\epsilon=0.54 \pm 0.07$. 
It is very interesting to notice 
that the size of the coherent source produced in $p\bar{p}$ collisions is around 
$2fm$ which is larger than 
the radius of chaotic source. Also from the above comparison we found that 
Eq.(21) is too simple to approximate Eq.(19).

Conclusions: In this paper, it was shown that the widely used two 
pion interferometry formula (Eq.(21)) was too simple to approximate Eq.(19). 
The information about the radius of coherent source  was lost in previous publication. 
Using Eq.(23) to fit the data, we found  that the radius of 
coherent source produced in $p\bar{p}$ collisions was around  $2 fm$. 
\acknowledgements

One of the author (Q.H.Z.) express his gratitude to Drs. U. Heinz, 
U. Widemann and Y. Pang  for helpful discussion.  
This work was supported by the Alexander 
von Humboldt Foundation and China national natural Science Foundation.

\end{document}